\documentclass[slac_one]{revtex4}
\usepackage{graphicx}
\usepackage{fancyhdr}
\def \bea{\begin{eqnarray}}
\def \beq{\begin{equation}}
\def \eeq{\end{equation}}
\def \lra{\Leftrightarrow}

\def \tu{\tau({\rm Universe})}

\begin{document}

\title{{\small{2005 ALCPG \& ILC Workshops - Snowmass,
U.S.A.}}\\ 
\vspace{12pt}
Dark Matter in Many Forms}

%

\author{Jonathan L. Rosner}
\affiliation{Enrico Fermi Institute, University of Chicago, 5640 S. Ellis
Avenue, Chicago, IL 60637}
\begin{abstract}

Since ordinary matter constitutes about 4\% of the closure density of the
Universe while dark matter constitutes about six times as much, it is
urged that searches for dark matter consider that it may exist in
several forms.  Implications for detection and hadron and $e^+ e^-$
colliders are discussed.

\end{abstract}

\maketitle

\thispagestyle{fancy}


\section{INTRODUCTION} 

Ordinary matter constitutes about 4\% of the closure density of
the Universe, while dark matter is responsible for about five times as much:
$\Omega_d = (23 \pm 4)\%$ \cite{Spergel:2003cb,Farrar:2005zd}.
Ordinary matter exists in several stable forms: $p$, $n$ (when incorporated
into nuclei), $e^-$, and three flavors of neutrinos.  (The lifetimes of the two
heavier mass eigenstates probably exceed the age of the Universe.)  We could
expect dark matter to exhibit at least as much variety \cite{variety}.

The observed space-time (4-dimensional) and rank (4) of the Standard Model
group SU(3) $\otimes$ SU(2) $\otimes$ U(1) is much less than the maximum number
of dimensions (10 or 11) considered in superstring theories or the rank of
typical groups (16) in such theories.  Moreover, there are at least two
well-motivated dark matter candidates already (axions and neutralinos), and
several variants of supersymmetry involve long-lived next-to-lightest
superpartners.  Thus it behooves us to case as wide a net as possible for
dark matter.  (The case for weakly interacting massive particles
(WIMPs) as dark matter is advanced forcefully in \cite{Battaglia:2005ie}.)

In the present report I give some motivations for multiple forms of
dark matter, discussing how several stable forms of ordinary matter arise
(Section II) and possible forms such variety might take for dark matter
(Section III).  Section IV is devoted to signatures in detectors planned for
the CERN Large Hadron Collider (LHC) and the International Linear Collider
(ILC).  It is not my intention to review signatures of all suggested dark
matter candidates, but to illustrate the broad range of possibilities.  For
reviews of dark matter candidates, see \cite{Bertone}.

\section{STABLE OBSERVED MATTER}

To describe the variety of stable forms of ordinary matter, I begin with the
simplest grand unified theory in which all fermions of a family belong to a
single representation, the group SO(10) \cite{SO10}.  The baryon number
$B$ and the lepton number $L$ are combined in a single charge $B-L$ conserved
in the SO(10) limit; quarks have $B-L = 1/3$ while leptons have $B-L = -1$.  No
separate labels exist for $B$ and $L$.  The existence of stable $qqq$
configurations is due to color SU(3).  Protons ($uud$) are long-lived in
comparison with $\tu$ as long as SO(10) gauge bosons mediating (e.g.) $ud \to
\bar d e^+$ are heavy enough.  Nonperturbative configurations
\cite{Kuzmin} enable $ud \lra \bar d e^+$ transitions but are only
operative at and above electroweak temperatures.

Free neutrons are unstable ($m_e + m_{\nu_e} + m_p < m_n$), but just barely.
They become stable when incorporated into some nuclei, leading to the richness
of ordinary matter.  The decay rates of the two heavier neutrino species in the
Standard Model should be of order $G_F^2 \alpha m_\nu^3 m_\ell^2/16 \pi^2$,
which is much larger than $\tu$.  One could not have anticipated three
quasi-stable neutrino species without understanding the existence of
quark-lepton families.  Neutrinos do contribute a non-dominant amount to the
dark matter of the Universe.

Thus, the variety of stable species in the Standard Model, including protons,
neutrons (in nuclei), electrons, and three families of neutrino, not to forget
the massless photon and graviton,
stems from a number of different sources.  It might be overly na\"{\i}ve to
expect stable dark matter to exist in only one form.

\section{OLD AND NEW QUANTUM NUMBERS}

Imagine a TeV-scale effective symmetry SU(3) $\otimes$ SU(2) $\otimes$ U(1)
$\otimes$ G, where G could be R-parity in supersymmetry, Kaluza-Klein parity
in theories with extra-dimensional excitations, T-parity in little Higgs models
\cite{LH}, Technicolor, or some other group.  We can classify the possible
types of matter from this standpoint as follows:
\begin{center}
\begin{tabular}{c c c c} \hline \hline
Type of matter & Std.\ Model &    G    & Example(s) \\ \hline
Ordinary       & Non-singlet & Singlet & Quarks, leptons \\
Mixed          & Non-singlet & Non-singlet & Superpartners \\
Shadow         & Singlet     & Non-singlet & $E_8'$ of E$_8 \otimes$
E$_8'$ \\ \hline \hline
\end{tabular}
\end{center}
Dark matter can take various forms, represented by each entry in the table.

\subsection{Ordinary Matter Examples}

Ordinary matter could be singlets under $G$ even if its subconstituents
were non-singlets.  In some composite-Higgs models, the gauge interaction
binding subconstituents implies additional baryon-like states.  Thus, ordinary
matter could indirectly already contain the hints of a structure which could
give rise to stable dark matter.  Ordinary matter could exist in unusual
configurations, corresponding to an alternative vacuum \cite{FN} or to
exotic states carrying baryon number \cite{Farrar:2005zd}.  These models
are claimed to account naturally for the ratio of dark to ordinary matter.

\subsection{Mixed Matter Examples}

Many dark matter scenarios involve mixed matter, such as superpartners or
particles with odd Kaluza-Klein- or T-parity.  These mixed-matter scenarios
may different from those conventionally discussed if G is more general
than a ``parity,'' for instance a nonabelian gauge group.
In supersymmetry there arises the possibility of non-topological solitons
known as ``Q-balls'' which have been proposed as dark matter candidates
\cite{QB}.

\subsection{Shadow Matter Examples}

Shadow matter may not interact with ordinary matter {\it at all} except
gravitationally.  Alternatively, shadow matter states may mix with those
of ordinary matter.  Examples may be found in the case of neutrinos.  For
instance, each SO(10) 16-dimensional spinor contains one right-handed neutrino
which is a singlet under the Standard Model group, in additional to
conventional quarks and leptons.  The grand unified group E$_{\rm 6}$, which
contains SO(10), has an additional sterile neutrino -- a singlet under
SO(10) -- in each 27-dimensional fundamental multiplet, in addition to
16- and 10-dimensional multiplets of SO(10).  Sterile neutrinos mixing with
ordinary ones have been proposed as dark matter candidates \cite{sn}.

\section{DETECTOR SIGNATURES}

Given the possibilities mentioned above, how can we be prepared for them?

Axion dark matter deserves increased attention.  In RF cavity searches a large
range of
frequencies remains to be scanned with adequate sensitivity to find axions even
if they were to account for {\it all} the expected dark matter \cite{ax}.  If
they constitute only {\it part} of it, the searches are even more challenging.
They become especially demanding
at higher frequencies (above the currently-studied range of up to 4 GHz,
corresponding to axions of mass greater than about $1.6 \times 10^{-5}$ eV),
since cavity design and tuning becomes progressively more difficult with
increasing frequency.

The neutrino dark matter contribution to $\Omega$ depends on
neutrinos' absolute masses, on which neutrinoless double beta
decay will shed some light.  We know that at least one neutrino species has a
mass exceeding about 0.05 eV.

Dark matter with non-singlet Standard Model charges but more than a Z$_2$
(parity) symmetry in the BSM group G may exist in several stable forms.  [In
the Standard Model, the color-singlet
$qqq$ configuration stems from the richness of the color SU(3) group.]
Even in supersymmetry there are scenarios in which the next-to-lightest
superpartner decays to the lightest superpartner over a non-prompt distance
(i.e., corresponding to a vertex displaced by at least a few tens of microns).
Detectors need to
be ready for kinks or vees with unexpected flight paths and for accumulation of
high-energy stable particles produced in pairs at high energies.  At the end
of experiments it may be worth searching these detectors for relics of
quasi-stable particles which have lodged in them and are still decaying
\cite{Feng}.

Another interesting signature of near-stability corresponds to intermittent
tracks in detectors.  One could imagine charged and neutral quasi-stable
particles split by so little in mass that they would repeatedly undergo
charge exchange with the detector and would leave a track looking like a dashed
line.

Dark matter with non-zero charges purely in the hidden sector will respond to
gravitational probes.  The detection of such particles in the mass range of
$10^{14}$ to $10^{20}$ gm has been discussed in Refs.\ \cite{Seto:2004zu}.

To summarize, exploring the full range of dark matter possibilities will
test our ingenuity!  I believe it is a prudent to base searches on
signatures suggested by the widest possible variety of theoretical frameworks.

\begin{acknowledgments}

This study was performed in part at the Aspen Center for Physics.
I wish to thank L. Dixon, G. Gelmini, J. Gunion, M. Peskin, J. Wells, and
L. Wolfenstein for helpful suggestions.  This work was supported in part by
the U. S. Department of Energy through Grant No.\ DE FG02 90ER40560.

\end{acknowledgments}

\end{document}